\documentclass{ecai}
\usepackage{times}
\usepackage{graphicx}
\usepackage{latexsym}
\usepackage{amsmath, amsthm, amssymb}
\usepackage{verbatim}
\usepackage{multicol, multirow}
\usepackage{booktabs}
\usepackage{caption}
\usepackage{float, graphicx}
\newtheorem{definition}{Definition}

\begin{document}
\title{A Novel Framework with Information Fusion and Neighborhood Enhancement for User Identity Linkage}

\author{Siyuan Chen \institute{School of Data and Computer Science, Sun Yat-sen University, China} \and Jiahai Wang \institute{School of Data and Computer Science, Sun Yat-sen University, China, email: wangjiah@mail.sysu.edu.cn. \it{$^*$Corresponding author: Jiahai Wang.}}$\ \ ^*$ \and Xin Du $^1$ \and Yanqing Hu $^1$}
\maketitle
\begin{abstract}
User identity linkage across social networks is an essential problem for cross-network data mining. Since network structure, profile and content information describe different aspects of users, it is critical to learn effective user representations that integrate heterogeneous information. This paper proposes a novel framework with INformation FUsion and Neighborhood Enhancement (INFUNE) for user identity linkage. The information fusion component adopts a group of encoders and decoders to fuse heterogeneous information and generate discriminative node embeddings for preliminary matching. Then, these embeddings are fed to the neighborhood enhancement component, a novel graph neural network, to produce adaptive neighborhood embeddings that reflect the overlapping degree of neighborhoods of varying candidate user pairs. The importance of node embeddings and neighborhood embeddings are weighted for final prediction. The proposed method is evaluated on real-world social network data. The experimental results show that INFUNE significantly outperforms existing state-of-the-art methods.

\end{abstract}
\section{Introduction}
Nowadays, users tend to have accounts in multiple social networks simultaneously for different services. For example, users may use LinkedIn to hunt for a job while adopting Instagram to share their daily life. Identifying linked accounts across different social networks enables us to integrate dispersed user information and provides a comprehensive understanding of user behavior, which will benefit many downstream implications, such as user profile modeling and cross-platform recommendation \cite{UILreview}. However, the correspondences among users’ different accounts, a.k.a. anchor links, are generally unavailable due to the independence of different social networks. Therefore, user identity linkage has become an increasingly popular area of research.

User information in social networks typically includes network structure (i.e., social connection), profile (e.g., screen name, location), and content (e.g., post), revealing different aspects of a single user. Some methods \cite{WhatIsInAName, NameBehavior, ULink, PALE, IONE, FRUI} use a single type of information. {A single type of information may be noisy and incomplete, and suffers from inconsistency across different social networks \cite{UILreview}}. Some methods \cite{MNA, PNA, HYDRA, COSNET, MASTER, Mego2Vec} explore the mechanisms to fuse multiple types of information. Multiple types of information may be complementary to each other.

Existing works focusing on information fusion can be divided into two categories, embedding based methods \cite{MASTER, Mego2Vec, LHNE, FactoidEmbedding} and non-embedding based methods \cite{MNA, PNA, HYDRA, COSNET}. Non-embedding based methods usually define hand-crafted features for every single type of information independently and combine them in a supervised or semi-supervised fashion \cite{MNA, HYDRA}. This paradigm fails to capture the deep semantic of user information and the rich interaction among different information. Unlike the former, embedding based methods \cite{DeepWalk, node2vec} seek to learn a low-dimensional vector representation for a user that preserves the characteristics of the original data. The vector representation can significantly reduce the cost of computation and storage, and can be easily incorporated with deep learning to obtain a flexible model.

Still, there remain two main problems for embedding based methods. First, none of them provide a general solution to fuse all three types of information, structure, profile and content. The embedding of network structure has been well-studied \cite{PALE, IONE, DALAUP}, and recently some works explore the joint embedding of structure and profile \cite{MASTER, Mego2Vec}, or structure and content \cite{LHNE}. Nevertheless, the latter requires specific interaction mechanisms between two types of information, which is not extensible to multiple types of information. Second, there have been few attempts to explicitly model the impact of the neighborhood, i.e., the first-order neighbors of a single user. Although matched neighbors have been utilized by non-embedding based methods, a typical embedding based model is only designed to learn node embeddings for node-level alignment, and is not designed to learn neighborhood embeddings for neighborhood-level alignment. Furthermore, as the matched neighbors vary with candidate user pairs, the neighborhood embeddings should be adapted dynamically, which is significantly different from the case of node embedding that seeks for a fixed representation for each user.

To tackle the challenges above, this paper proposes a novel framework with INformation FUsion and Neighborhood Enhancement (INFUNE) for user identity linkage. INFUNE contains two components, an information fusion component and a neighborhood enhancement component. The information fusion component employs a {group of encoders and decoders to preserve the characteristic of each type of information, and integrate them in the node embeddings.} Based on the node embeddings, the potential matched neighbors of a given user pair can be identified, and the neighborhood enhancement component, a novel graph neural network model, is applied to learn adaptive neighborhood representations.

The main contributions of this work are summarized as follows.
\begin{itemize}
\item An information fusion component is proposed to integrate different user information in a unified manner. To our best knowledge, this is the first attempt to fuse user information of structure, profile and content simultaneously for user identity linkage in an embedding based model.
\item To utilize the potential matched neighbors for user identity linkage, INFUNE employs a novel graph neural network to learn neighborhood representations that vary with candidate user pairs.
\item Extensive experiments are conducted to validate the performance of INFUNE. The results show the superiority of INFUNE by comparing with state-of-the-art models.
\end{itemize}
\section{Related Work}
As there are multiple types of user information including network structure, profile and content, the existing works can be divided into two main categories, one exploits only a single type of user information while the other aims to integrate multiple types of information.

For the first category of methods, linking users by comparing attributes of profiles is most widely studied \cite{WhatIsInAName, NameBehavior, ULink}. Mu et al. \cite{ULink} map multiple attributes to a common latent space, where matched users lie closer than unmatched ones. Apart from the profile, users' generated content can be utilized to extract more features \cite{HYDRA, MNA}. Kong et al. \cite{MNA} convert the posts of each user into a bag-of-words vector weighted by TF-IDF and calculated the cosine similarities. However, profile and content information are generally incomplete and inconsistent, while network structure is more accessible and more consistent across social networks. Many methods \cite{PALE, IONE, GraphUIL} adopt network embedding to encode the structure information to low-dimensional vectors and predicted linked users via vector similarities. Nevertheless, the sparsity of the network structure prevents these methods to learn discriminative user representations.

To overcome the drawbacks of every single type of information, many researchers seek to combine them, leading to the second category of methods. Zhong et al. \cite{CoLink} create independent models for profile and structure, and makes them reinforce each other iteratively using a co-training algorithm. Su et al. \cite{MASTER} map users into a latent space that preserves both structure and profile similarities. Zhang et al. \cite{Mego2Vec} learn the profile embeddings from character level and word level, and aggregate the information of neighbors using an attention mechanism. To combine structure and content information, Wang et al. \cite{LHNE} extracted topics from content and defined a user-topic network to learn unified user embeddings. Li et al. \cite{UUIL, SNNA, MSUIL} adopt TADW \cite{TADW} to fuse structure and content information, and link users by aligning the distributions of social networks with known anchor links as learning guidance. However, the aforementioned methods are designed to integrate only two types of information and the well-designed interaction mechanisms between heterogeneous information can hardly extend to integrate multiple types of information.

Among all the methods mentioned above, \cite{PALE, IONE, MASTER, Mego2Vec, GraphUIL, LHNE, UUIL, SNNA, MSUIL} are embedding based methods and \cite{WhatIsInAName, NameBehavior, ULink, MNA, HYDRA, CoLink} are non-embedding based methods.
\section{Problem Formulation}

Let $\mathcal{G} = (\mathcal{U}, \mathcal{E}, \mathcal{P}, \mathcal{C})$ denote a social network, where $\mathcal{U}=\{u_i\}^N_{i=1}$ is the set of users, $\mathcal{E}\subseteq \mathcal{U} \times \mathcal{U}$ is the set of social connections, $\mathcal{P}$ is the set of user profiles, and $\mathcal{C}$ is the set of user generated contents. Each user $u \in \mathcal{U}$ is associated with a profile $p \in \mathcal{P}$ and a content $c \in \mathcal{C}$. Each $p$ contains several user attributes, such as screen name, location and description, depending on the dataset. Each $c$ contains a set of texts generated by a single user.

This paper focuses on the problem of linking users between two social networks. Without loss of generality, one is regarded as the source network, while the other is regarded as the target network, denoted as $\mathcal{G}^s$ and $\mathcal{G}^t$, respectively. The problem of user identity linkage is defined as follows.
\begin{definition}[User Identity Linkage]
Given two social networks $\mathcal{G}^s$ and $\mathcal{G}^t$, the task of user identity linkage is to find a function $\mathcal{F}: \mathcal{U}^s \times \mathcal{U}^t \to \{0, 1\}$ such that $\forall (u_i, u_j) \in \mathcal{U}^s \times \mathcal{U}^t$,
\begin{equation}
\mathcal{F}\left(u_i, u_j\right)=
\begin{cases}
1, & \textnormal{if } u_i \textnormal{ and } u_j \textnormal{ belongs to the same person},\\
0, & \textnormal{otherwise}.
\end{cases}
\end{equation}

\end{definition}
From the definition above, it is sufficient to assess the pairwise similarities among users and generate potential matched user pairs by ranking the similarities. The similarity can be evaluated from different perspectives since users possess several types of raw features. However, the information from different similarity indicators can be redundant or contradictory. For example, an individual may maintain similar friends and keep similar writing styles in multiple social networks, while using completely different screen names for privacy protection. This results in the agreement of structural similarity and content similarity, and their disagreement with profile similarity. Therefore, it is challenging to unify different similarity indicators for user identity linkage. A naive solution is to train a binary classifier with similarity vectors as the input, while this method fails to capture the complex relations among different similarity indicators. Recently, Hamilton et al. \cite{LeskovecReview} point out that various network embedding models can be unified in an encoder-decoder model that reconstructs the pairwise similarities among nodes within a graph. This paper extends this framework to learn user embeddings that preserve multiple types of similarities simultaneously, and the unified similarities are evaluated based on the embeddings for the final task.
\section{Proposed Method}
\begin{figure}[!t]
\centering
\includegraphics[width=0.9\linewidth]{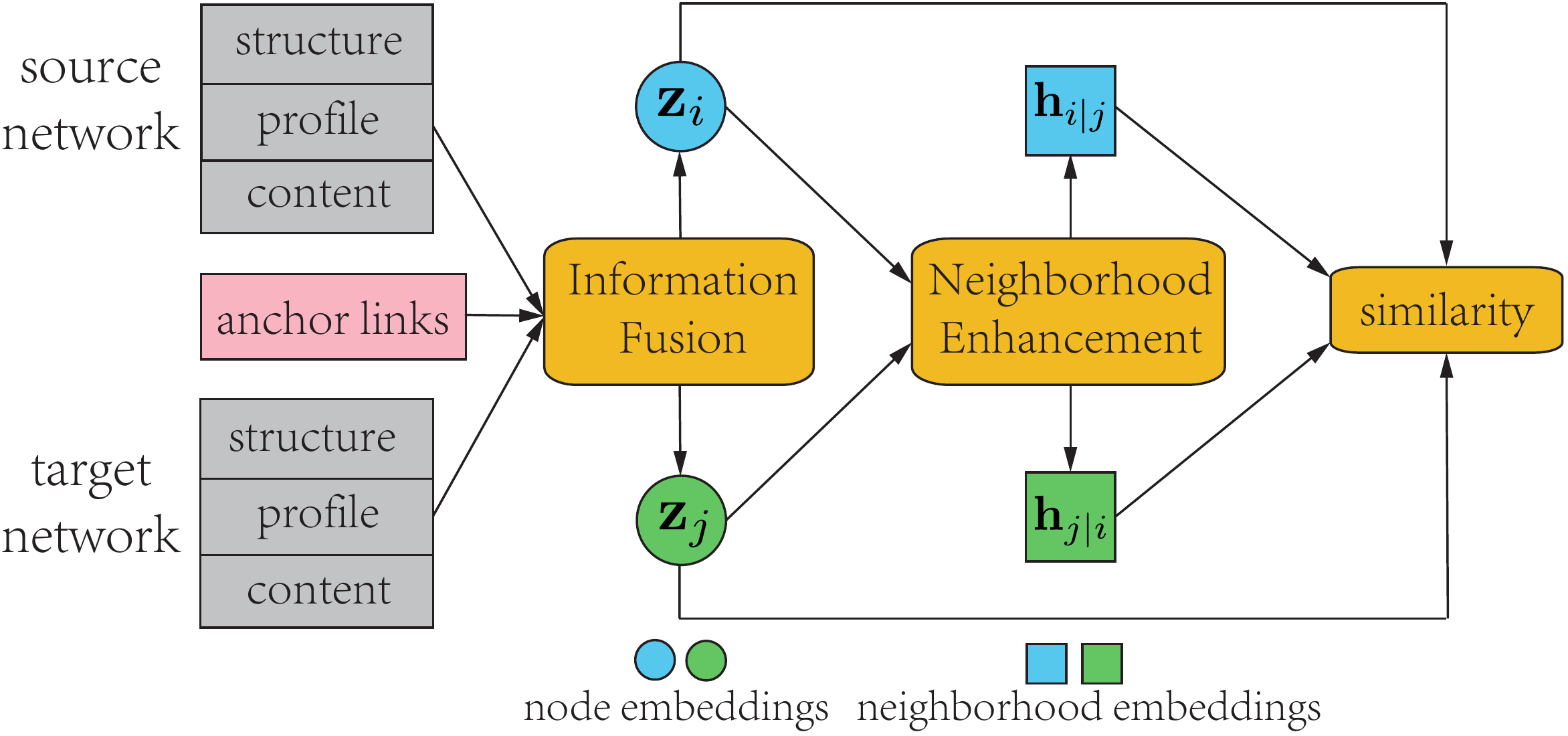}
\caption{Overview of INFUNE. User information and supervised information are fed to the information fusion component to produce node embeddings $\mathbf{z}_i$ and $\mathbf{z}_j$ for users from the source network and the target network, respectively. $\mathbf{z}_i$ and $\mathbf{z}_j$ are fed to the neighborhood enhancement component to generate the corresponding neighborhood embeddings $\mathbf{h}_{i|j}$ and $\mathbf{h}_{j|i}$.}
\label{overview}
\end{figure}
The structure of INFUNE is presented in Figure \ref{overview}. INFUNE contains two components, the information fusion component and the neighborhood enhancement component. The raw features of users, including structure, profile and content, together with known anchor links, are first preprocessed as different similarity matrices. The resulting similarity matrices are fed into the information fusion component to obtain the node embeddings with heterogeneous information. The node embeddings are ready for preliminary comparison. Based on the node embeddings, the neighborhood enhancement component first identifies potential matched neighbors of candidate user pairs and then learns adaptive neighborhood embeddings that reflect the overlapping degree of the neighborhoods of candidate user pairs. Finally, a weighted sum of node similarity and neighborhood similarity is evaluated as the unified similarity for user identity linkage.

\subsection{Information Fusion Component}
A simple scheme for information fusion is to learn the embeddings of a user for different features independently and unify them into a single vector. This usually requires to design a sophisticated collaboration mechanism, since simple methods like concatenation fail to capture the complex interactions among features. Worse still, the number of parameters grows linearly with the numbers of features and users, which is not scalable to large social networks with heterogeneous information. To address the problems above, an information fusion component is proposed as follows.
\subsubsection{Component Overview}
\begin{figure}[h]
\centering
\includegraphics[width=0.85\linewidth]{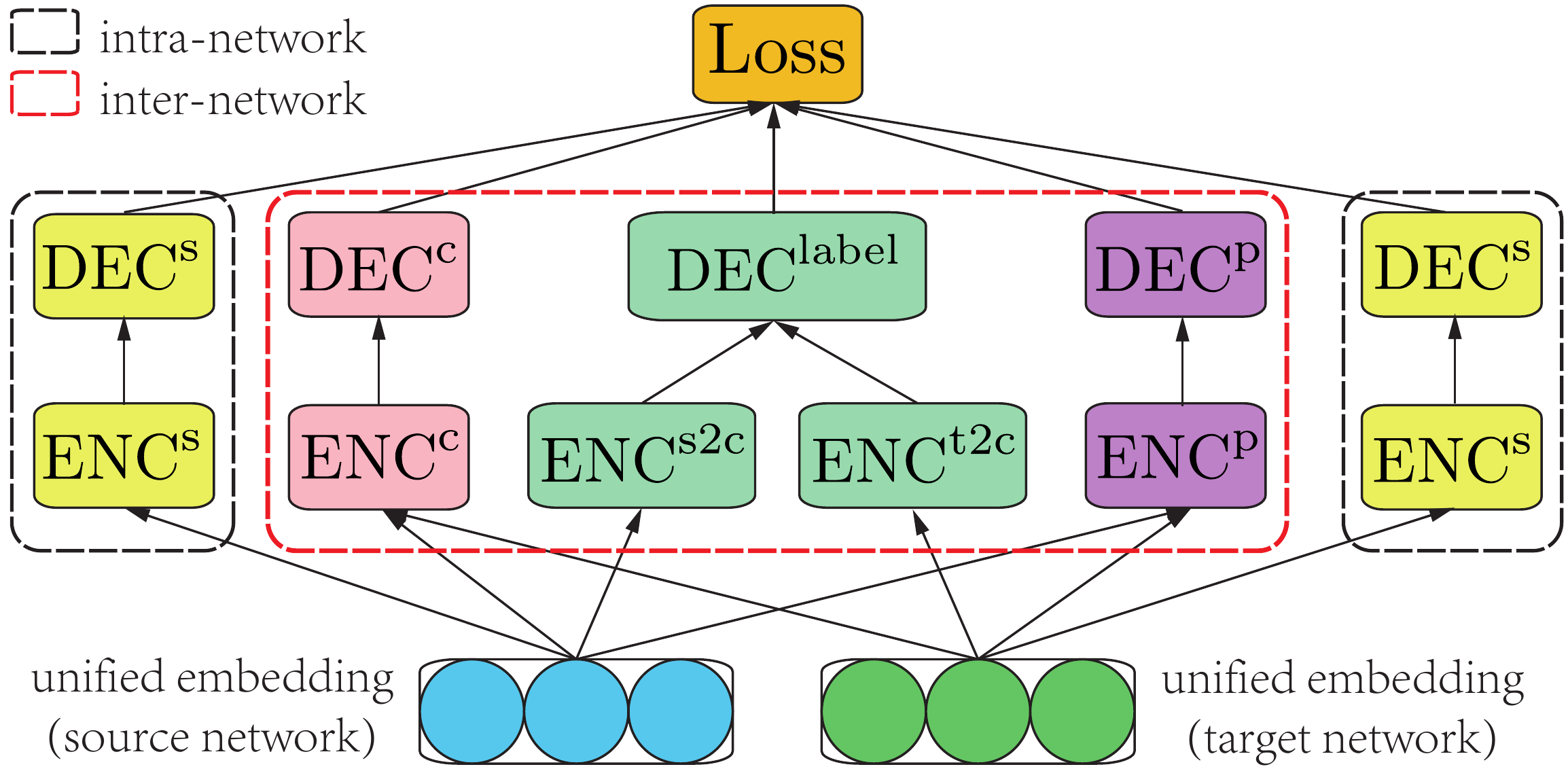}
\caption{Information fusion component. A group of feature-specific encoders (ENC) and decoders (DEC) is adopted to preserve user similarities w.r.t. different raw features. Both intra-network and inter-network similarities are considered in this component.}
\label{ENC-DEC}
\end{figure}
As shown in Figure \ref{ENC-DEC}, the information fusion component assigns each user a vector, named the unified embedding, to integrate multiple types of information in a single social network. The unified embeddings are fed to different pairs of encoders and decoders to preserve user similarities w.r.t. different information, and all pairs work similarly. By leveraging known anchor links as the supervised information, the unified embeddings of users from different social networks are mapped to a common latent space. The resulting embeddings are called the node embeddings. The details of the information fusion component are described as follows.

Let $\mathbf{x} \in \mathbb{R}^D$ be the unified embedding of some user, where $D \ll N$ is the dimension of the embedding. $\mathbf{x}$ is mapped to different feature spaces through different feature-specific encoders to preserve the characteristic of the corresponding raw features. Formally, $\forall \alpha \in \{\text{s}, \text{p}, \text{c}\}$, short for $\{\text{structure}, \text{profile}, \text{content}\}$, the feature embedding of $\mathbf{x}$ is defined by
\begin{equation}\label{encoder}
\mathbf{z}^{\alpha} = \text{ENC}^{\alpha}(\mathbf{x}) = \mathbf{W}_2^{\alpha} \tanh \left(\mathbf{W}_1^{\alpha} \mathbf{x} + \mathbf{b}_1^{\alpha} \right) + \mathbf{b}_2^{\alpha},
\end{equation}
where $\text{ENC}^{\alpha}$ can be any learnable linear or non-linear mapping, and for simplicity, a two-layer perceptron is adopted. The usage of encoders avoids explicitly maintaining embedding matrices for different features, which greatly reduces the number of parameters.

For any two users, $u_i \in \mathcal{U}^1$ and $u_j \in \mathcal{U}^2$, a feature-specific similarity indicator is defined as follows,
\begin{equation}
g^{\alpha}_{ij} = \text{sim}^{\alpha}\left(u_i, u_j\right),
\end{equation}
where $g^{\alpha}_{ij}$ is called the ground truth similarity between $u_i$ and $u_j$ w.r.t. feature $\alpha$.
Note that $\text{sim}^{\alpha}$ can be intra-network ($\mathcal{U}^1 = \mathcal{U}^2$) or inter-network ($\mathcal{U}^1 \neq \mathcal{U}^2$), and can be symmetric or asymmetric, depending on the features (Cf. Section \ref{structure} and Section \ref{profile and content}).
Correspondingly, a feature-specific decoder is designed to reconstruct the user similarity between $u_i$ and $u_j$ w.r.t. feature $\alpha$, i.e.,
\begin{equation}\label{rec_sim}
r^{\alpha}_{ij} = \text{DEC}^{\alpha}\left(\mathbf{z}^{\alpha}_i, \mathbf{z}^{\alpha}_j\right),
\end{equation}
where $\text{DEC}^{\alpha}$ can be some operator like inner product or cosine similarity, or a learnable module, and $r^{\alpha}_{ij}$ is called the reconstructed similarity between $u_i$ and $u_j$ w.r.t. feature $\alpha$.

Notably, different from most of the existing network embedding models that apply to a single graph, the embeddings of users from two social networks are passed through the same encoders and decoders, which can help achieve alignment in different feature spaces.

The discrepancy between the reconstructed similarity $r^{\alpha}_{ij}$ and the true value $g^{\alpha}_{ij}$ is measured by a loss function $\ell^{\alpha} \left(r^{\alpha}_{ij}, g^{\alpha}_{ij}\right)$, and the empirical loss $\mathcal{L}^{\alpha}$ over all user pairs is
\begin{equation}
\mathcal{L}^{\alpha} = \frac{1}{N_1 N_2} \sum_{u_i \in \mathcal{U}^1} \sum_{u_j \in \mathcal{U}^2} \ell^{\alpha} \left(r^{\alpha}_{ij}, g^{\alpha}_{ij}\right),
\end{equation}
where $N_1$ and $N_2$ are the number of users in $\mathcal{U}^1$ and that of $\mathcal{U}^2$, respectively.

In this paper, $\ell^{\alpha}$ is chosen to be the squared loss for all $\alpha$. Let $\mathbf{R}^{\alpha}, \mathbf{G}^{\alpha} \in \mathbb{R}^{N_1 \times N_2}$ be the reconstructed similarity matrix and the ground truth similarity matrix, respectively, then the objective $\mathcal{L}^{\alpha}$ can be rewritten in a compact matrix form,
\begin{equation}
\mathcal{L}^{\alpha} = \frac{1}{N_1 N_2} \left\| \mathbf{R}^{\alpha} - \mathbf{G}^{\alpha} \right\|^2_F,
\end{equation}
where $\| \cdot \|_F$ is the Frobenius norm.

The formulations above are not only designed for fusing the information of the raw features, but they can also be applied to incorporate the supervised information by constructing a binary matrix that indicates whether or not two users are matched (Cf. Section \ref{supervised information}). Denoting the corresponding loss as $\mathcal{L}^{\text{label}}$, the overall objective for information fusion is
\begin{equation}
\label{all_loss}
\mathcal{L}^{\text{all}} = \mathcal{L}^{\text{label}} + \sum_{\alpha \in \{\text{s}, \text{p}, \text{c}\}} \mathcal{L}^{\alpha}.
\end{equation}
\subsubsection{Structure Embedding}\label{structure}Asymmetric relations, e.g., follower-followee relations, ubiquitously exist in social networks. Regarding them as symmetric relations will fail to capture these features that are useful for user identity linkage \cite{IONE}. Therefore, for the structure information, it is intuitive to define an intra-network similarity indicator that tells if there is a directed edge between two users. The similarity indicator is asymmetric, which requires to design an asymmetric decoder. The key is to model a directed edge. Most of the existing structural embedding methods \cite{IONE} define two embeddings for a node, one as the source node embedding and the other as the target node embedding. However, this requires an extra embedding for a node and fails to model the connection between the two roles of a node in a network. Note that social connection can be regarded as a separate object that is independent of the nodes it links, and it is shared by the two social networks. This observation inspires us to explicitly modeling a directed edge by defining a transformation $\varphi$ that maps the structural embedding $\mathbf{z}^{\text{s}}$, regarding as the source node embedding, to a target node embedding $\mathbf{z}^{\text{t}} = \varphi (\mathbf{z}^{\text{s}})$.
For simplicity, $\varphi$ is chosen to be a two-layer perceptron as in Eq. (\ref{encoder}).

Based on the transformation $\varphi$, an asymmetric decoder can be defined as follows,
\begin{equation}\label{truncated cosine}
\text{DEC}^{\text{s}}\left(\mathbf{z}^{\text{s}}_i, \mathbf{z}^{\text{s}}_j\right)
= \text{cos}^+\left(\mathbf{z}^{\text{s}}_i, \mathbf{z}^{\text{t}}_j\right)
\triangleq \max\left\{ 0, \text{cos}\left<\mathbf{z}^{\text{s}}_i, \mathbf{z}^{\text{t}}_j\right>\right\},
\end{equation}
where the cosine similarity is adopted to measure the linking strength between two nodes, and the negative values are truncated to restrict the range of the reconstructed similarities to $[0, 1]$.

\subsubsection{Profile and Content Embedding}\label{profile and content}
Inter-network similarity indicators are adopted for profile and content information, since there can be several shared attributes of profiles between two social networks, and contents can be compared directly via many text similarity indicators. Screen names are selected to measure the profile similarity, as they have been proved to be effective in user identity linkage \cite{WhatIsInAName}. The normalized Levenstein distance is used to compute the string similarities among screen names. To compute the content similarity, first all posts of a user are concatenated to a single document, and then documents of all users from two social networks are fed to a Doc2Vec model \cite{doc2vec} to obtain the text embeddings, and finally, as in Eq. (\ref{truncated cosine}), the truncated cosine similarity is used to measure the similarities among the text embeddings.

Naturally, symmetric decoders for profile and content embeddings can be defined as follows,
\begin{equation}
\text{DEC}^{\alpha}\left(\mathbf{z}^{\alpha}_i, \mathbf{z}^{\alpha}_j\right) = \text{cos}^+\left(\mathbf{z}^{\alpha}_i, \mathbf{z}^{\alpha}_j\right),
\alpha \in \{\text{p}, \text{c}\}.
\end{equation}
\subsubsection{Supervised Information} \label{supervised information}

The unified embeddings of users from different social networks can not be compared directly as they lie in different vector spaces. Therefore, two encoders, $\text{ENC}^{\text{s2c}}$ and $\text{ENC}^{\text{t2c}}$ are introduced for the source network and the target network respectively to map different users to a common latent space, i.e.,
\begin{equation}
\label{st2c}
\begin{split}
\mathbf{z}_i &= \text{ENC}^{\text{s2c}}\left( \mathbf{x}_i\right),\\
\mathbf{z}_j &= \text{ENC}^{\text{t2c}}\left( \mathbf{x}_j\right),
\end{split}
\end{equation}
where $\mathbf{x}_i$ and $\mathbf{x}_j$ represent the unified embeddings of $u_i \in \mathcal{U}^{\text{s}}$ and $u_j \in \mathcal{U}^{\text{t}}$, repectively.

Intuitively, $\text{sim}^{\text{label}}$ can be defined as an indicator function that indicates if two users are matched. Similar to the decoders for profile and content, a symmetric decoder for the supervised information can be defined as follows,
\begin{equation}
\label{eq:dec_node}
r^{\text{node}}_{ij} = \text{DEC}^{\text{label}}\left(\mathbf{z}_i, \mathbf{z}_j\right) = \text{cos}^+\left(\mathbf{z}_i, \mathbf{z}_j\right).
\end{equation}
\subsubsection{Optimization}\label{optimization}
Since the overall objective $\mathcal{L}^{\text{all}}$ is the sum of similar loss functions for different information, it is sufficient to consider the optimization of a single loss function. To simplify the formulation, the superscript $\alpha$ for all related symbols are omitted.

The key challenge for optimization is the sparsity of the ground truth matrix
$\mathbf{G}$. Generally, only a small fraction of users are highly similar to a given user, while the rest are dissimilar. Regarding the similar and dissimilar users as positive samples and negative samples, respectively, directly optimizing $\mathcal{L}$ tends to overfit on negative samples and underfit on positive samples, which prevents the models from learning discriminative embeddings of users for user identity linkage. Besides, the time complexity is $O\left( N_1 N_2\right)$, which is costly for large-scale social networks. Inspired by Mikolov et al. \cite{word2vec}, a negative sampling trick is introduced to address the problems above.

Formally, given a percentage $\theta \in [0, 1]$, for any $u_i \in \mathcal{U}^{1}$, the $\theta$-quantile of the $i$-th row of $\mathbf{G}$ is denoted as $q^{\theta}_i$, and $\mathcal{U}^{2}$ is split into two disjoint subsets,
\begin{equation}
\begin{split}
\mathcal{U}^2_+ (i) &= \left\{ u_j \in \mathcal{U}^2
\,\middle|\, g_{ij} \geq q^{\theta}_i \right\}, \\
\mathcal{U}^2_- (i) &= \left\{ u_j \in \mathcal{U}^2
\,\middle|\, g_{ij} < q^{\theta}_i\right\},
\end{split}
\end{equation}
representing the sets of similar users and dissimilar users, respectively. Then, $\mathcal{L}$ can be reformulated as follows,
\begin{equation}
\label{opt_loss}
\resizebox{1.0\columnwidth}{!}{$
\begin{split}
\mathcal{L} &= \frac{1}{N_1 N_2} \left\| \mathbf{R} - \mathbf{G} \right\|^2_F \\
\mathcal{L} &= \frac{1}{N_1 N_2} \sum_{u_i \in \mathcal{U}^1} \sum_{u_j \in \mathcal{U}^2} \left(r_{ij} - g_{ij}\right)^2 \\
&= \frac{1}{N_1 N_2} \sum_{u_i \in \mathcal{U}^1} \left\{ \sum_{u_j \in \mathcal{U}^2_+(i)} \left(r_{ij} - g_{ij}\right)^2 + \sum_{u_j \in \mathcal{U}^2_-(i)} \left(r_{ij} - g_{ij}\right)^2 \right\} \\
& \approx \frac{1}{M} \sum_{u_i \in \mathcal{U}^1} \sum_{u_j \in \mathcal{U}^2_+(i)} \left\{\left(r_{ij} - g_{ij}\right)^2 + \sum^K_{n=1} \underset{u_n \sim P(j)}{\mathbb{E}} \left[\left(r_{in} - g_{in}\right)^2\right] \right\} \\
&\triangleq \widehat{\mathcal{L}}
\end{split}
$}
\end{equation}
where $M$ is a normalization constant, $K$ is the number of negative samples $u_n$ sampled from the ``noisy distribution" $P(j) \propto d_j^{3/4}$, and $d_j = \sum^{N_1}_{i=1} |g_{ij}|$. It can be calculated that $M = (K + 1)\sum^{N_1}_{i=1} \left| \mathcal{U}^2_+(i)\right|$. Therefore, the time complexity of evaluating the loss function is reduced to $O(M)$. For a large percentage $\theta$ and a small number $K$, the evaluation is more efficient than that of the original formulation. Specifically, if $\mathbf{G}$ is an adjacency matrix, $O(M) = O \left(K \left| \mathcal{E}\right| \right)$, which coincides with the time complexity of most existing structural embedding models \cite{LINE}. $\widehat{\mathcal{L}}$ is used as the final loss function, and all parameters are updated via gradient descent.
\subsection{Neighborhood Enhancement Component}
The node embedding $\mathbf{z}$ obtained by the information fusion component can be directly applied for user identity linkage. However, this matching scheme ignores the effect of common neighbors, which can result in mistakes that the neighbors of the predicted matched users are mostly unmatched. To promote the precision of user identity linkage, a neighborhood enhancement component is applied to learn neighborhood embeddings that reflect the overlapping degree of neighborhoods of candidate user pairs.

\begin{figure}[h]
\centering
\includegraphics[width=0.85\linewidth]{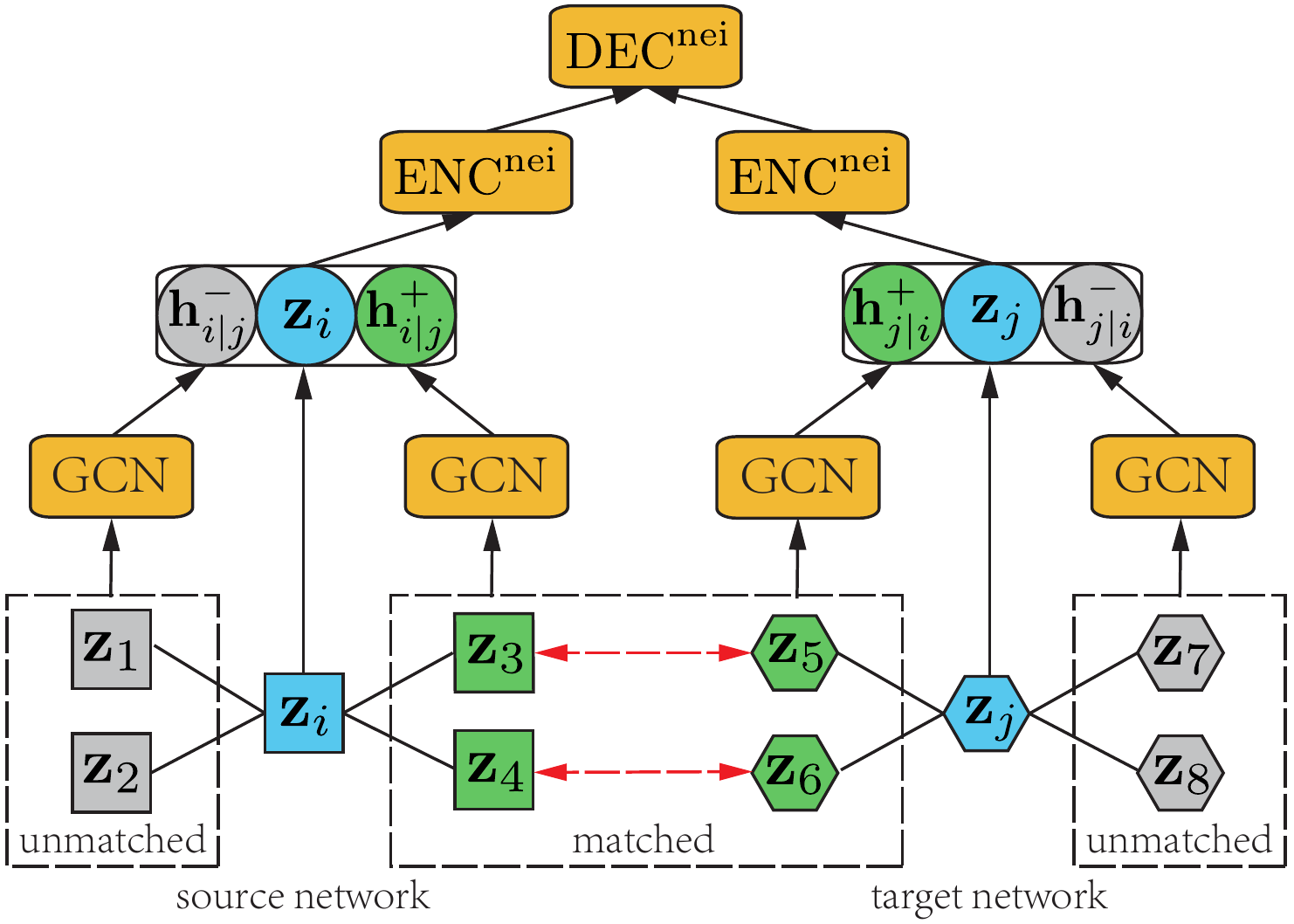}
\caption{Neighborhood enhancement component. $\mathbf{z}_i$ and $\mathbf{z}_j$ represent the node embeddings of $u_i \in \mathcal{U}^{\text{s}}$ and $u_j \in \mathcal{U}^{\text{t}}$, respectively, while $\left\{\mathbf{z}_n\right\}^8_{n=1}$ represent the embeddings of their neighbors. For $u_i$, the node embeddings of the potential matched and unmatched neighbors are fed to a graph convolution network (GCN) to obtain $\textbf{h}^+_{i|j}$ and $\textbf{h}^-_{i|j}$, the neighborhood embeddings of the potential matched and unmatched neighbors, respectively. The concatenation of $\textbf{z}_i$, $\textbf{h}^+_{i|j}$ and $\textbf{h}^-_{i|j}$ are fed to an encoder ($\text{ENC}^{\text{nei}}$) to obtain the neighborhood embedding of $u_i$. The same procedure can be applied to $u_j$ to obtain its neighborhood embedding. The similarity of the two neighborhood embeddings is measured by a decoder ($\text{DEC}^{\text{nei}}$).}
\label{PAGCN}
\end{figure}

An intuitive solution is to use a graph convolution network (GCN) \cite{GCN} to aggregate the information of neighbors. However, GCN convolves the embeddings of matched neighbors and unmatched neighbors indiscriminately, which may bring in additional noise that hurts the precision of matching. Besides, the obtained neighborhood embeddings are fixed, while the common neighbors vary with candidate user pairs. To address the problems above, the neighborhood enhancement component, shown in Figure \ref{PAGCN}, aggregates the information of the matched and unmatched neighbors separately, and unify them to the neighborhood embeddings, which are adaptive with varying candidate user pairs.
First, given a user pair, potential matched neighbors are identified with the ``$\text{one-to-one}_{\leq}$" constraint \cite{PNA}, i.e., each user from the source network can be mapped to at most one user in the target network. Then, the neighborhood is split into two disjoint subsets, containing potential matched neighbors ($+$) and unmatched neighbors ($-$), respectively. Formally, for $u_i \in \mathcal{U}^{\text{s}}$ and $u_j \in \mathcal{U}^{\text{t}}$,
\begin{equation}
\begin{split}
\mathcal{N}_i &= \mathcal{N}^+_{i|j} + \mathcal{N}^-_{i|j}, \\
\mathcal{N}_j &= \mathcal{N}^+_{j|i} + \mathcal{N}^-_{j|i}, \\
\left| \mathcal{N}^+_{i|j}\right| &= \left| \mathcal{N}^+_{j|i}\right|,
\end{split}
\end{equation}
where $\mathcal{N}_i$ and $\mathcal{N}_j$ stand for the sets of neighbors of $u_i$ and $u_j$, respectively.

Second, GCN is applied to obtain the neighborhood embeddings of potential matched neighbors and unmatched neighbors. For $u_i$, the two embeddings are defined by
\begin{equation}
\begin{split}
\mathbf{h}^+_{i|j} &= \text{GCN} \left(\mathcal{N}^+_{i|j}\right) = \frac{1}{\left| \mathcal{N}^+_{i|j}\right|} \sum_{u_n \in \mathcal{N}^+_{i|j}} \mathbf{z}_n, \\
\mathbf{h}^-_{i|j} &= \text{GCN} \left(\mathcal{N}^-_{i|j}\right) = \frac{1}{\left| \mathcal{N}^-_{i|j}\right|} \sum_{u_n \in \mathcal{N}^-_{i|j}} \mathbf{z}_n.
\end{split}
\end{equation}
Then, $\mathbf{h}^+_{i|j}$ and $\mathbf{h}^-_{i|j}$ are concatenated with $\mathbf{z}_i$ to obtain an embedding that integrates the information of $u_i$ and $\mathcal{N}_i$. Finally, this embedding is fed to a two-layer perceptron to obtain the neighborhood embedding, i.e.,
\begin{equation}
\label{emb_nei}
\mathbf{h}_{i|j} = \text{ENC}^{\text{nei}} \left( \mathbf{z}_i \oplus \mathbf{h}^+_{i|j} \oplus \mathbf{h}^-_{i|j}\right).
\end{equation}
A similar procedure can be applied to $u_j$ to obtain $\mathbf{h}_{j|i}$.

The ground truth similarity is defined as in Section~\ref{supervised information}, and the reconstructed similarity between $\mathcal{N}_i$ and $\mathcal{N}_j$ is defined as
\begin{equation}
\label{nei_sim}
r^{\text{nei}}_{ij} = \text{DEC}^{\text{nei}}\left(\mathbf{h}_{i|j}, \mathbf{h}_{j|i}\right) = \text{cos}^+\left(\mathbf{h}_{i|j}, \mathbf{h}_{j|i}\right).
\end{equation}

The loss function of the neighborhood enhancement component is defined as in Eq.~(\ref{opt_loss}) and the parameters can be updated via gradient descent.

With the node embeddings and the neighborhood embeddings, the total similarity between $u_i$ and $u_j$ is defined to be a weighted sum of the node similarity and the neighborhood similarity,
\begin{equation}
\label{total_sim}
r_{ij}^{\text{total}} = \frac{1}{1 + \lambda}\left(r_{ij}^{\text{node}} + \lambda \cdot r_{ij}^{\text{nei}}\right), \lambda \geq 0,
\end{equation}
where $\lambda$ is a tunable hyperparameter that measures the importance of the neighborhood similarity.

\section{Experimental Evaluation}
\subsection{Dataset and Experimental Settings}
\paragraph{Dataset.} This paper uses a dataset collected from two Chinese social networks, Douban (https://www.douban.com) and Weibo (https://www.weibo.com). Douban is a Chinese social networking service website that allows users to record information and create content related to films, books, etc. There have been 200 million registered users as of 2013. Weibo is a leading micro-blogging platform in China, with over 445 million monthly active users as of 2018. Users create original content or retweet as on Twitter.

This dataset contains network structure, profile and content information. An anchor link is constructed if there is a Weibo homepage address link in the profile of the Douban user. Compared with existing public datasets, our dataset contains many more users, resulting in more anchor links and richer user relationships. Besides, our dataset contains a large number of contents, which are not included in existing public datasets. These characteristics pose more challenges to user identity linkage task. The statistics are listed in Table \ref{dataset}.

\begin{table}
\centering
\caption{Statistics of the Douban-Weibo Dataset.}
\resizebox{0.99\columnwidth}{!}{
\begin{tabular}{cccccc}
\toprule
Network & \#Users & \#Edges & \#Posts & \#Anchors & \#Shared Edges \\
\midrule
Douban & 9734 & 260467 & 10941957 & \multirow{2}[4]{*}{9514} & \multirow{2}[4]{*}{55207} \\
\cmidrule{1-4} Weibo & 9514 & 196978 & 3799357 & & \\
\bottomrule
\end{tabular}
}
\label{dataset}
\end{table}

The follower-followee relations are regarded as directed edges. Screen names are used as the profile information, and the missing values are imputed by empty strings. For content information, LTP \cite{LTP}, a Chinese language processing toolkit, is used for word segmentation. All posts of a user are merged to a single document and the bag-of-words model is used for text preprocessing.

\paragraph{Parameter Settings.} The dimensions of user embeddings of all methods are set to be $256$. In our method, the numbers of hidden neurons of all encoders are set to be $512$. For the negative sampling procedure, the percentage $\theta$ is set to be $0.99$, and the number of negative samples $K$ is set to be $5$. For the neighborhood enhancement component, only the top-$250$ similar users are selected to evaluate neighborhood similarities, and the weight $\lambda$ is set to be $0.2$ according to the grid searching results and a detailed discussion is shown in Section~\ref{NEM}.

All methods are evaluated at different ratios of the training set. The ratio $\eta$ ranges from $0.1$ to $0.9$.

\paragraph{Metric.} The hit-precision is selected as the evaluation metric to compare the top-$k$ candidates, which is well-established and widely-used in many real user linkage applications \cite{ULink}. This paper sets $k=30$ and evaluates all competitive methods by computing the top-$k$ precision for each test user as follows,
\begin{equation}
h(u) =
\begin{cases}
\frac{k - (\text{hit}(u) - 1)}{k}, &\text{if}\ k \geq \text{hit}(u) \geq 1, \\
0, &\text{otherwise.}
\end{cases}
\end{equation}
where $\text{hit}(u)$ represents the position of the correctly identified user in the returned top-$k$ users. Then, the hit-precision is calculated on $N$ test users by $\frac{1}{N} \sum^N_{i=1} h(u_i)$.
\subsection{Baselines}
To evaluate the performance of INFUNE, we compare it with several state-of-the-art methods listed as follows.
\begin{itemize}
\item ULink \cite{ULink}: a non-embedding based method that projects the raw feature vectors of user profiles to a common latent space.

\item PALE \cite{PALE}: an embedding based method that first embeds the network structure to a low-dimensional space, and then learns a mapping function in a supervised manner.

\item GraphUIL \cite{GraphUIL}: an embedding based method that applies a graph neural network to jointly capture local and global network structure information.

\item MASTER \cite{MASTER}: an embedding based method that maps users into a latent space that preserves intra-network structure and profile similarities.

\item MEgo2Vec \cite{Mego2Vec}: an embedding based method that learns profile embeddings from character level and word level, and aggregates the information of potential matched neighbors using attention mechanism.

\end{itemize}

The codes of ULink\footnote{http://www.lamda.nju.edu.cn/code\_ULink.ashx} and MEgo2Vec\footnote{https://github.com/BoChen-Daniel/MEgo2Vec-Embedding-Matched-Ego-Networks-for-User-Alignment-Across-Social-Networks} are public and thus directly used in our experiments. Other baselines are implemented by ourselves according to the original papers. The code of INFUNE\footnote{https://github.com/hilbert9221/INFUNE} is available online. To verify the effectiveness of the information fusion component and the neighborhood component, some variants of INFUNE are introduced as follows.
\begin{itemize}
\item $\text{INFUNE}_{\text{s}}$, $\text{INFUNE}_{\text{p}}$ and $\text{INFUNE}_{\text{c}}$: variants of INFUNE using a single type of information, i.e., network structure, profile or content information.
\item $\text{INFUNE}_{\text{sp}}$, $\text{INFUNE}_{\text{sc}}$ and $\text{INFUNE}_{\text{pc}}$: variants of INFUNE using the pairwise combinations of the three types of information.
\item $\text{INFUNE}_{-\text{NE}}$: a variant of INFUNE without the neighborhood enhancement component.
\end{itemize}

Besides, the user representations of some baselines such as ULink
and GraphUIL can be easily replaced by the unified embeddings generated by an unsupervised version of INFUNE. To further verify the effectiveness of the information fusion component, this paper feeds the unified embeddings to ULink and GraphUIL, and the resulting variants are denoted as $\text{ULink}_\text{spc}$ and $\text{GraphUIL}_\text{spc}$, respectively.

\subsection{Comparisons with Baselines}
\begin{figure}[h]
\centering
\includegraphics[width=0.85\linewidth]{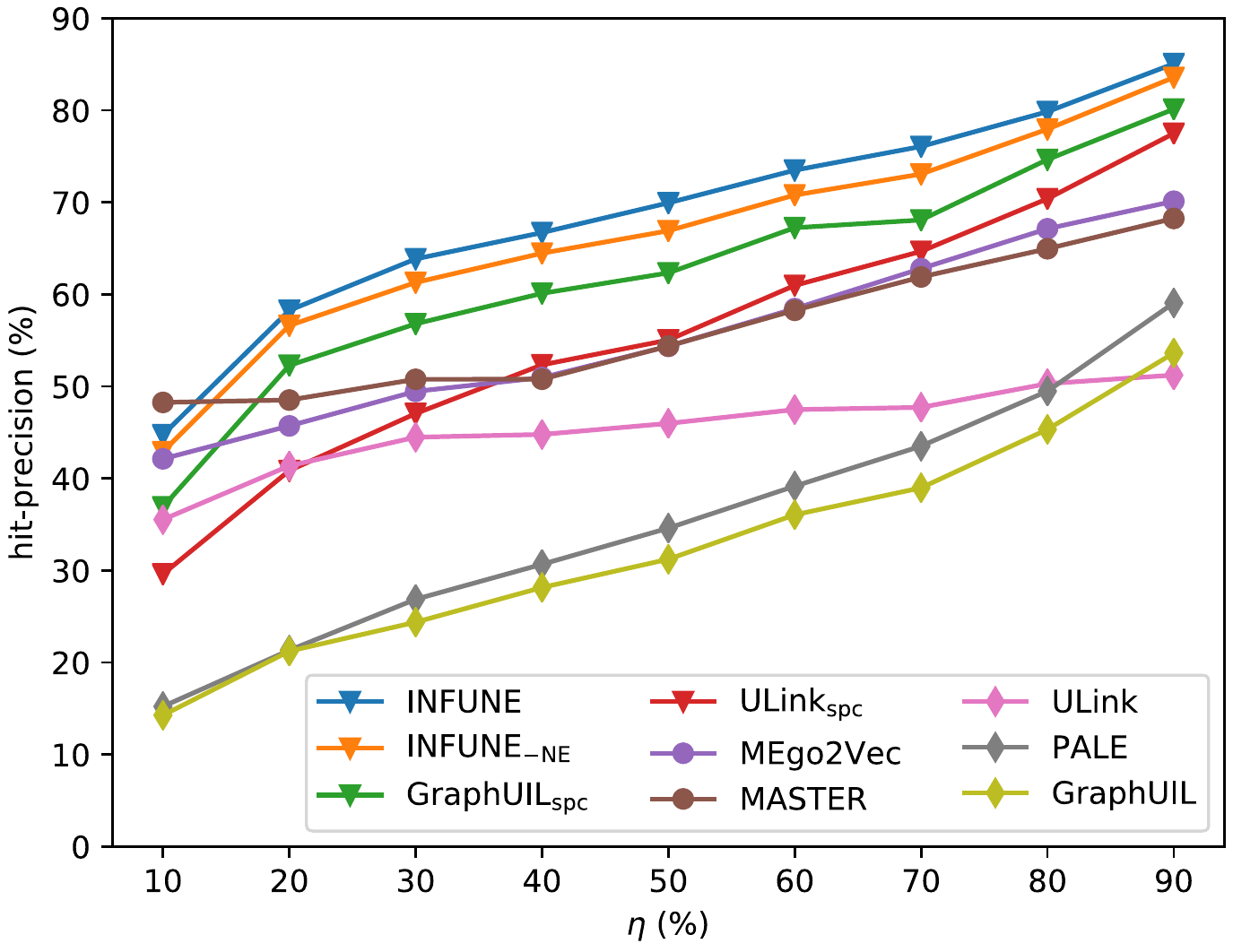}
\caption{Comparisons with baselines.}
\label{exp-algorithms}
\end{figure}
Figure \ref{exp-algorithms} shows the overall performance of all methods on the Douban-Weibo dataset. The proposed method performs clearly better than the baselines (+12.43\% on average). Our model achieves higher hit-precision than MEgo2Vec and MASTER, probably because they ignore the content information. Besides, compared with MEgo2Vec, maybe our model considers the effect of not only potential matched neighbors but also that of unmatched neighbors. Compared with MASTER, maybe our model utilizes not only intra-network but also inter-network similarities to learn user embeddings. Note that our model performs a bit worse than MASTER at $\eta = 10\%$. Maybe INFUNE additionally fuses content information, requiring more training data to better map users from different social networks to a common space. With larger $\eta$, INFUNE significantly outperforms MASTER, indicating that INFUNE can better leverage supervised information. ULink, PALE and GraphUIL perform the worst as they rely on a single type of information that suffers from inconsistency across social networks.

\subsection{Effect of Information Fusion}
\begin{figure}[h]
\centering
\includegraphics[width=0.85\linewidth]{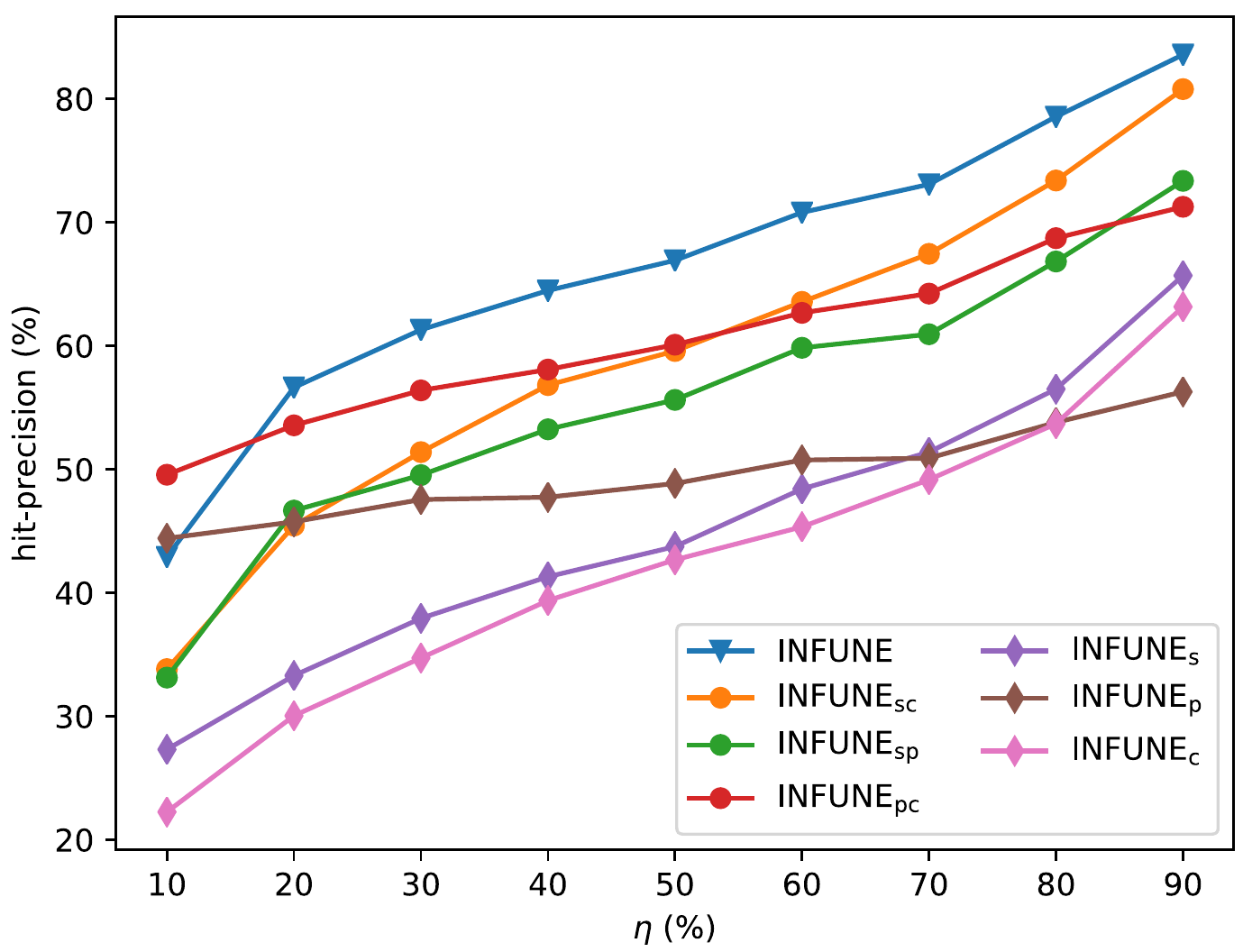}
\caption{Comparisons with variants of INFUNE.}
\label{exp-ablations}
\end{figure}
Variants removing a single type of information and two types of information are compared with INFUNE to verify the effectiveness of information fusion, and the results are shown in Figure~\ref{exp-ablations}. Compared with INFUNE, the hit-precisions of $\text{INFUNE}_{\text{sp}}$, $\text{INFUNE}_{\text{sc}}$ and $\text{INFUNE}_{\text{pc}}$ decrease by 13.21\%, 9.54\% and 8.17\% on average, respectively, which demonstrates that each single type of information is indispensable for user identity linkage. Among all information, the content information contributes the most according to the decrease of the hit-precision, which may be because content contains richer information than the others. The importance of structure and profile information depends on the ratio of the training set. When $\eta$ is larger than $0.5$, the structure information contributes more than the profile information, while the situation is reversed when $\eta$ is smaller than $0.5$. Besides, the performance of INFUNE is even worse than $\text{INFUNE}_{\text{pc}}$ at $\eta = 0.1$. The results indicate that the alignment of network structure relies more on the known anchor links, while the alignment based on the profile information is less sensitive to the supervised information.

By removing two types of information, the hit-precisions decrease by 19.12-26.41\%, which again verifies the effectiveness of information fusion. Judging from the decrease of the hit-precision, the combination of structure and content is the most effective with sufficient supervised information ($\eta > 0.7$). When $\eta \leq 0.7$, the combination of structure and profile outperforms the others. The performance of $\text{INFUNE}_{\text{s}}$ is close to that of $\text{INFUNE}_{\text{c}}$, which indicates that the effect of the combination of profile and content is similar to that of the combination of structure and profile. Interestingly, the performance of $\text{INFUNE}_{\text{p}}$ is more stable than INFUNE with varying $\eta$, which suggests that linking users based on the combination of structure and content requires more supervised information.

As shown in Figure 4, $\text{ULink}_\text{spc}$ and $\text{GraphUIL}_\text{spc}$ significantly outperform ULink and GraphUIL, respectively, which again verifies that embeddings with multiple types of information are more effective than those with a single type of information. Still, INFUNE performs better than $\text{ULink}_\text{spc}$ and $\text{GraphUIL}_\text{spc}$, which indicates that our whole framework with two integrated components can better leverage heterogeneous information.

\subsection{Effect of Neighborhood Enhancement}
\label{NEM}
\begin{figure}[h]
\centering
\includegraphics[width=0.85\linewidth]{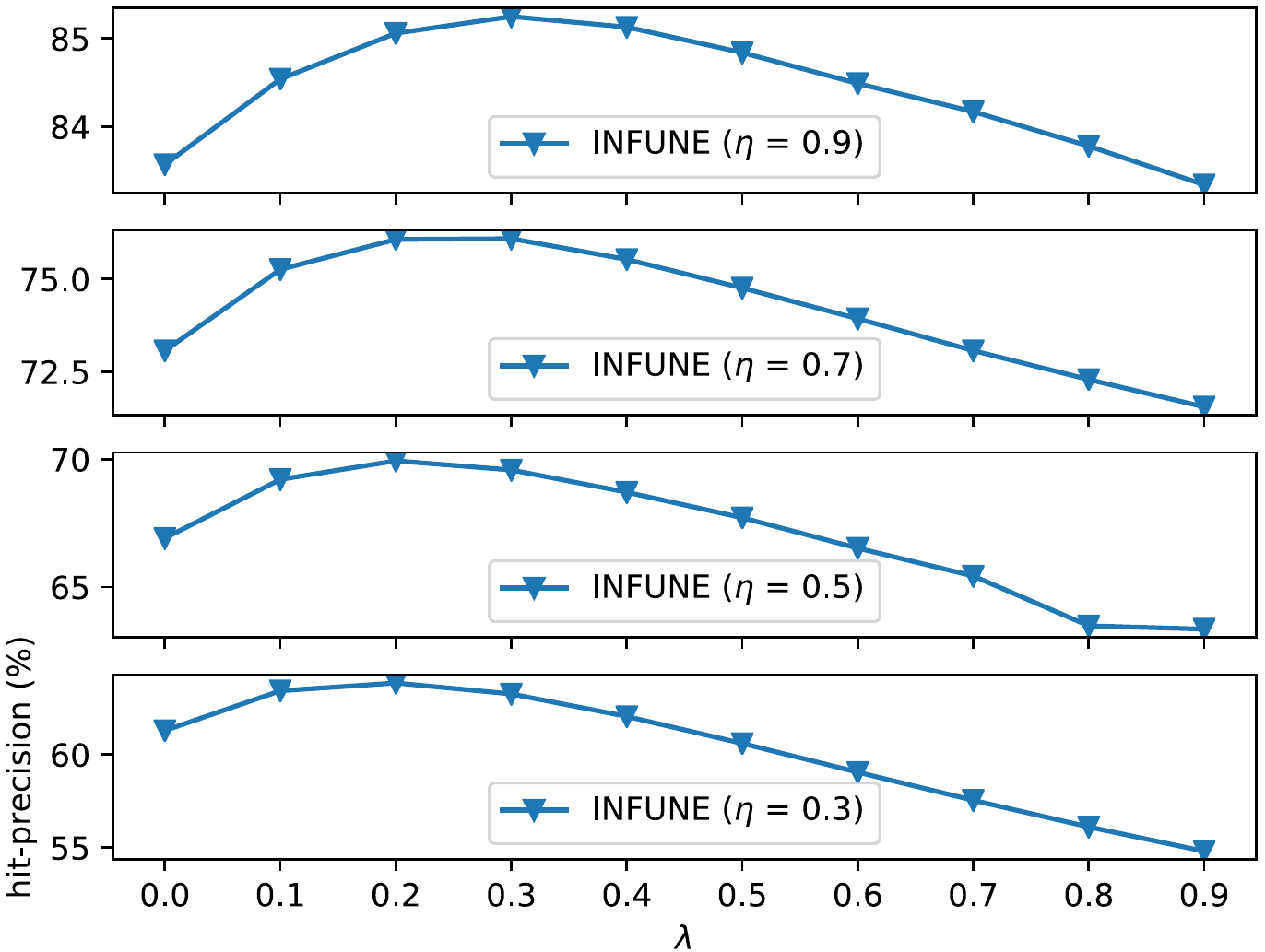}
\caption{Performance of INFUNE w.r.t. $\lambda$.}
\label{exp-pagcn}
\end{figure}
As shown in Figure~\ref{exp-algorithms}, by removing the neighborhood enhancement component, the performance decreases by 2.20\% on average, but the difference is less significant for $\eta < 0.3$ and $\eta > 0.7$. Possible reasons are as follows. When $\eta$ is smaller than $0.3$, the known common neighbors are sparse, and therefore, they are inadequate to help identify matched users. With sufficient supervised information, i.e. when $\eta > 0.7$, node similarities dominate the results of user identity linkage and the neighborhood information is less helpful.

The parameter $\lambda$ plays an important role in leveraging neighborhoods for user identity linkage. $\lambda$ is chosen from $\{0, 0.1,\dots, 0.9\}$. When $\lambda = 0$, INFUNE is reduced to $\text{INFUNE}_{-\text{NE}}$. Figure \ref{exp-pagcn} shows the effect of $\lambda$ at selected ratios of the training set, i.e., $\eta \in \{ 0.3, 0.5, 0.7, 0.9 \}$. It is observed that despite different $\eta$, the hit-precision first increases with $\lambda$, peaks at around $\lambda = 0.2$, and finally decreases as
$\lambda$ grows. When $\lambda$ is larger than $0.4$, the performance of INFUNE tends to be worse than $\text{INFUNE}_{-\text{NE}}$, that is, large $\lambda$ may bring in a negative effect to the whole model. The results indicate that node similarities dominate the performance of the whole model, and neighborhood similarities with a small weight can improve the hit-precision by 1.49-3.03 \%. This is in line with our common sense that dissimilar users may share many common neighbors in social networks, and therefore, common neighbors contribute less to identify matched users.

\subsection{Visualization of Learned Embeddings}
\begin{figure}[h]
\centering
\includegraphics[width=0.9\linewidth]{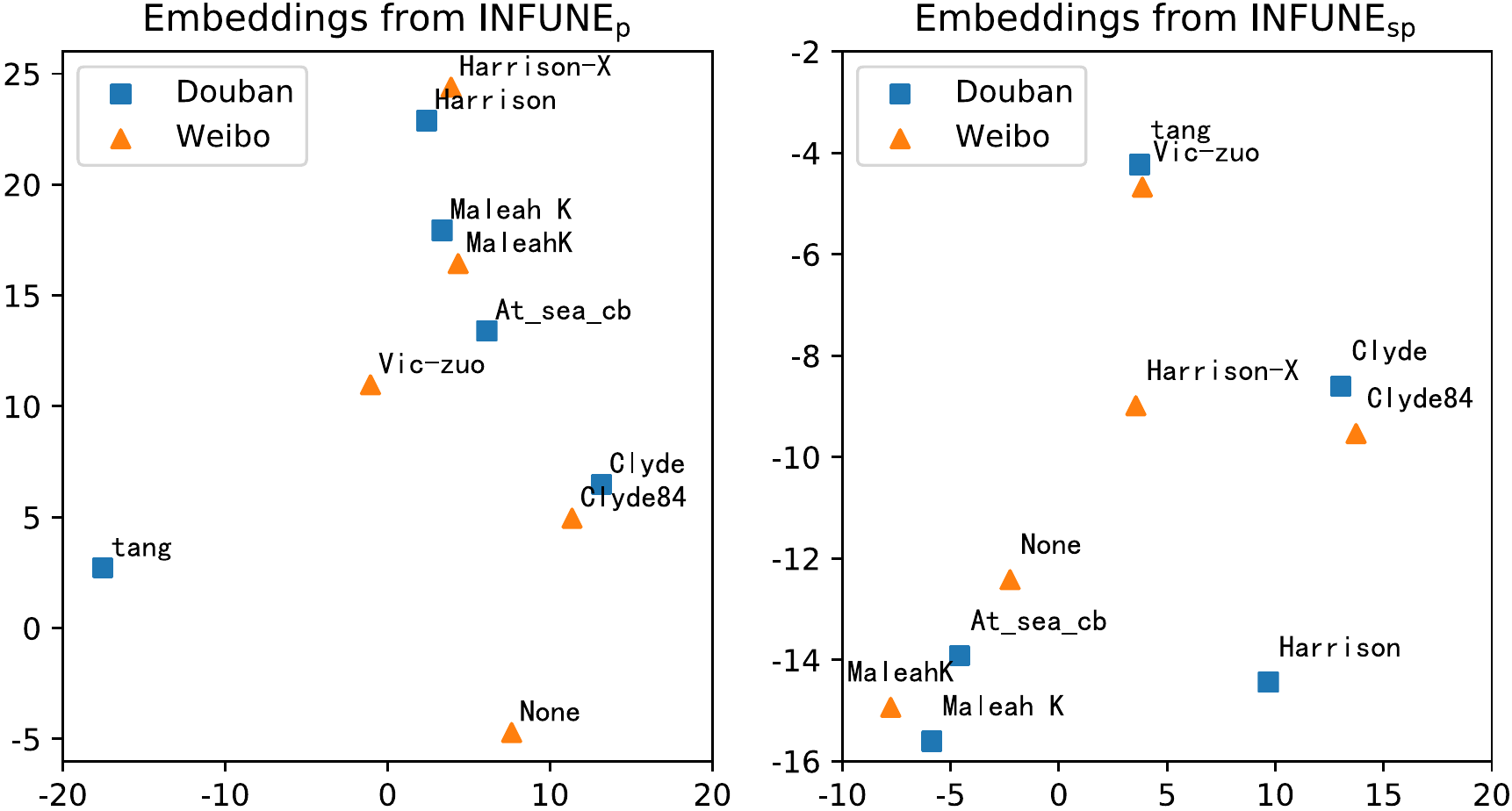}
\caption{Visualization of learned embeddings. (Maleah K, MaleahK), (Clyde, Clyde84), (tang, Vic-zuo) and (At\_sea\_cb, None) are matched user pairs, where None stands for a missing screen name. (Harrison, Harrison-X) is an unmatched user pair.}
\label{psp}
\end{figure}
A dimension reduction algorithm t-SNE \cite{t-SNE} is adopted to project the node embeddings to two-dimensional space to illustrate the complementarity of heterogeneous information. As users often publish hundreds of posts, it is hard to show the content consistency of matched users within limited space. For simplicity and clarity, the node embeddings generated by $\text{INFUNE}_{\text{p}}$ and $\text{INFUNE}_{\text{sp}}$ are selected for visualization. Figure \ref{psp} visualizes the node embeddings of four randomly selected matched user pairs and one unmatched user pair. The vector spaces of the two types of embeddings are referred to as the profile space and the structure-profile space, respectively. The square points represent users from Douban, and triangle points represent the users from Weibo. Note that the embeddings of the same user can be completely different as they lie in different vector spaces, and thus, it is the relative position between users that matters.

User pairs with similar screen names lie closer to each other than the others in the profile space. Based on the profile information, $\text{INFUNE}_{\text{p}}$ correctly identifies the matched user pairs (Maleah K, MaleahK) and (Clyde, Clyde84), but mistakenly matches user pair (Harrison, Harrison-X). Worse still, $\text{INFUNE}_{\text{p}}$ fails to identify matched user pairs (tang, Vic-zuo) and (At\_sea\_cb, None) with completely different screen names or even missing screen names. Thus, the prediction of $\text{INFUNE}_{\text{p}}$ is not reliable and robust for users with noisy and incomplete profile information.

By adding the structure information, $\text{INFUNE}_{\text{sp}}$ inherits the advantages of $\text{INFUNE}_{\text{p}}$ and eliminates its limitations over the given user pairs. First of all, the relative positions of the identified matched user pairs are consistent with those in the profile space, which means that they will not be incorrectly predicted as unmatched user pairs. Secondly, unmatched users Harrison and Harrison-X with similar screen names are separated from each other in the structure-profile space, which shows that structure information can help distinguish similar users from each other. Finally, matched user pairs with dissimilar screen names, i.e., (tang, Vic-zuo) and (At\_sea\_cb, None) are successfully clustered, which again demonstrates the complementarity of profile and structure information.

\section{Conclusion and Future Work}
This paper presents a novel framework with information fusion and neighborhood enhancement for user identity linkage. The information fusion component effectively learns node embeddings that integrate user information of structure, profile and content by reconstructing pairwise user similarities. The neighborhood enhancement component adopts a novel graph neural network to learn adaptive neighborhood embeddings that reflect the overlapping degree of the neighborhoods of candidate user pairs. The results of extensive experiments on real-world social network data validate the effectiveness of our method.

Currently, embedding mechanisms for heterogeneous contents, e.g., texts and images, are not yet considered as our dataset contains only textual content. A solution is to learn content embeddings separately by reconstructing intra-network similarities and to map them to a common space by utilizing the supervised information. Exploration for more sophisticated methods will be left for future work. Besides, the two components of our model are trained separately, as the neighborhood enhancement component relies on the discriminative node embeddings generated by a well-trained information fusion component. Future work includes jointly training the two components to promote performance.

\ack This work is supported by the National Key R\&D Program of China (2018AAA0101203), and the National Natural Science Foundation of China (61673403, U1611262).
\bibliographystyle{ecai}
{\tiny
\bibliography{ref}

\begin{thebibliography}{10}

\bibitem{LTP}
Wanxiang Che, Zhenghua Li, and Ting Liu, `Ltp: A chinese language technology
  platform', in {\em COLING}, pp. 13--16, (2010).

\bibitem{DALAUP}
Anfeng Cheng, Chuan Zhou, Hong Yang, Jia Wu, Lei Li, Jianlong Tan, and Li~Guo,
  `Deep active learning for anchor user prediction', in {\em IJCAI}, pp.
  2151--2157, (2019).

\bibitem{node2vec}
Aditya Grover and Jure Leskovec, `node2vec: Scalable feature learning for
  networks', in {\em KDD}, pp. 855--864. ACM, (2016).

\bibitem{LeskovecReview}
William~L Hamilton, Rex Ying, and Jure Leskovec, `Representation learning on
  graphs: Methods and applications', {\em arXiv preprint arXiv:1709.05584},
  (2017).

\bibitem{GCN}
Thomas~N. Kipf and Max Welling, `Semi-supervised classification with graph
  convolutional networks', in {\em ICLR}, (2017).

\bibitem{MNA}
Xiangnan Kong, Jiawei Zhang, and Philip~S Yu, `Inferring anchor links across
  multiple heterogeneous social networks', in {\em CIKM}, pp. 179--188. ACM,
  (2013).

\bibitem{doc2vec}
Quoc Le and Tomas Mikolov, `Distributed representations of sentences and
  documents', in {\em ICML}, pp. 1188--1196, (2014).

\bibitem{MSUIL}
Chaozhuo Li, Senzhang Wang, Hao Wang, Yanbo Liang, Philip~S Yu, Zhoujun Li, and
  Wei Wang, `Partially shared adversarial learning for semi-supervised
  multi-platform user identity linkage', in {\em CIKM}, pp. 249--258, (2019).

\bibitem{SNNA}
Chaozhuo Li, Senzhang Wang, Yukun Wang, Philip Yu, Yanbo Liang, Yun Liu, and
  Zhoujun Li, `Adversarial learning for weakly-supervised social network
  alignment', in {\em AAAI}, volume~33, pp. 996--1003, (2019).

\bibitem{UUIL}
Chaozhuo Li, Senzhang Wang, Philip~S Yu, Lei Zheng, Xiaoming Zhang, Zhoujun Li,
  and Yanbo Liang, `Distribution distance minimization for unsupervised user
  identity linkage', in {\em CIKM}, pp. 447--456. ACM, (2018).

\bibitem{WhatIsInAName}
Jing Liu, Fan Zhang, Xinying Song, Young-In Song, Chin-Yew Lin, and Hsiao-Wuen
  Hon, `What's in a name?: an unsupervised approach to link users across
  communities', in {\em WSDM}, pp. 495--504. ACM, (2013).

\bibitem{IONE}
Li~Liu, William~K Cheung, Xin Li, and Lejian Liao, `Aligning users across
  social networks using network embedding.', in {\em IJCAI}, pp. 1774--1780,
  (2016).

\bibitem{HYDRA}
Siyuan Liu, Shuhui Wang, Feida Zhu, Jinbo Zhang, and Ramayya Krishnan, `Hydra:
  Large-scale social identity linkage via heterogeneous behavior modeling', in
  {\em SIGMOD}, pp. 51--62. ACM, (2014).

\bibitem{t-SNE}
Laurens van~der Maaten and Geoffrey Hinton, `Visualizing data using t-sne',
  {\em JMLR}, {\bf 9}(Nov),  2579--2605, (2008).

\bibitem{PALE}
Tong Man, Huawei Shen, Shenghua Liu, Xiaolong Jin, and Xueqi Cheng, `Predict
  anchor links across social networks via an embedding approach', in {\em
  IJCAI}, volume~16, pp. 1823--1829, (2016).

\bibitem{word2vec}
Tomas Mikolov, Kai Chen, Greg Corrado, and Jeffrey Dean, `Efficient estimation
  of word representations in vector space', in {\em ICLR Workshop}, (2013).

\bibitem{ULink}
Xin Mu, Feida Zhu, Ee-Peng Lim, Jing Xiao, Jianzong Wang, and Zhi-Hua Zhou,
  `User identity linkage by latent user space modelling', in {\em KDD}, pp.
  1775--1784. ACM, (2016).

\bibitem{DeepWalk}
Bryan Perozzi, Rami Al-Rfou, and Steven Skiena, `Deepwalk: Online learning of
  social representations', in {\em KDD}, pp. 701--710. ACM, (2014).

\bibitem{UILreview}
Kai Shu, Suhang Wang, Jiliang Tang, Reza Zafarani, and Huan Liu, `User identity
  linkage across online social networks: A review', {\em ACM SIGKDD
  Explorations Newsletter}, {\bf 18}(2),  5--17, (2017).

\bibitem{MASTER}
Sen Su, Li~Sun, Zhongbao Zhang, Gen Li, and Jielun Qu, `Master: across multiple
  social networks, integrate attribute and structure embedding for
  reconciliation.', in {\em IJCAI}, pp. 3863--3869, (2018).

\bibitem{LINE}
Jian Tang, Meng Qu, Mingzhe Wang, Ming Zhang, Jun Yan, and Qiaozhu Mei, `Line:
  Large-scale information network embedding', in {\em WWW}, pp. 1067--1077,
  (2015).

\bibitem{LHNE}
Yaqing Wang, Chunyan Feng, Ling Chen, Hongzhi Yin, Caili Guo, and Yunfei Chu,
  `User identity linkage across social networks via linked heterogeneous
  network embedding', {\em WWW},  1--22, (2018).

\bibitem{FactoidEmbedding}
Wei Xie, Xin Mu, Roy Ka-Wei Lee, Feida Zhu, and Ee-Peng Lim, `Unsupervised user
  identity linkage via factoid embedding', in {\em ICDM}, pp. 1338--1343. IEEE,
  (2018).

\bibitem{TADW}
Cheng Yang, Zhiyuan Liu, Deli Zhao, Maosong Sun, and Edward Chang, `Network
  representation learning with rich text information', in {\em IJCAI}, (2015).

\bibitem{NameBehavior}
Reza Zafarani and Huan Liu, `Connecting users across social media sites: a
  behavioral-modeling approach', in {\em KDD}, pp. 41--49. ACM, (2013).

\bibitem{PNA}
Jiawei Zhang, Weixiang Shao, Senzhang Wang, Xiangnan Kong, and Philip~S Yu,
  `Partial network alignment with anchor meta path and truncated generic stable
  matching', {\em arXiv preprint arXiv:1506.05164}, (2015).

\bibitem{Mego2Vec}
Jing Zhang, Bo~Chen, Xianming Wang, Hong Chen, Cuiping Li, Fengmei Jin, Guojie
  Song, and Yutao Zhang, `Mego2vec: embedding matched ego networks for user
  alignment across social networks', in {\em CIKM}, pp. 327--336. ACM, (2018).

\bibitem{GraphUIL}
Wen Zhang, Kai Shu, Huan Liu, and Yalin Wang, `Graph neural networks for user
  identity linkage', {\em arXiv preprint arXiv:1903.02174}, (2019).

\bibitem{COSNET}
Yutao Zhang, Jie Tang, Zhilin Yang, Jian Pei, and Philip~S Yu, `Cosnet:
  Connecting heterogeneous social networks with local and global consistency',
  in {\em KDD}, pp. 1485--1494. ACM, (2015).

\bibitem{CoLink}
Zexuan Zhong, Yong Cao, Mu~Guo, and Zaiqing Nie, `Colink: An unsupervised
  framework for user identity linkage', in {\em AAAI}, (2018).

\bibitem{FRUI}
Xiaoping Zhou, Xun Liang, Haiyan Zhang, and Yuefeng Ma, `Cross-platform
  identification of anonymous identical users in multiple social media
  networks', {\em IEEE TKDE}, {\bf 28}(2),  411--424, (2015).

\end{thebibliography}
}
\end{document}